\documentclass[pdflatex,sn-aps]{sn-jnl}

\usepackage{graphicx}%
\usepackage{multirow}%
\usepackage{amsmath,amssymb,amsfonts}%
\usepackage{amsthm}%
\usepackage{mathrsfs}%
\usepackage[title]{appendix}%
\usepackage{xcolor}%
\usepackage{textcomp}%
\usepackage{manyfoot}%
\usepackage{booktabs}%
\usepackage{algorithm}%
\usepackage{algorithmicx}%
\usepackage{algpseudocode}%
\usepackage{listings}%

\theoremstyle{thmstyleone}%
\theoremstyle{thmstyletwo}%

\theoremstyle{thmstylethree}%

\begin{document}

\title[Packaging particles in curved spacetimes]{Packaging particles in curved spacetimes}


\author*[1]{\fnm{Jesse} \sur{Huhtala}}\email{jejohuh@utu.fi}
\author[1]{\fnm{Iiro} \sur{Vilja}}\email{vilja@utu.fi}

\affil[1]{\orgdiv{Department of Physics and Astronomy}, \orgname{University of Turku}, \orgaddress{\city{20014 Turku},  \country{Finland}}}


\abstract{Scattering calculations in curved spacetime are technically complicated and, in the case of a general spacetime metric, quite impossible. Even in the cases where perturbative scattering calculations can be done one has to be careful about what kind of particles are sensible to measure. Curved spacetime quantum field theories are then less conceptually clear than those in flat spacetime. In this article, we investigate an aspect of this conceptual confusion -- the use of wave packets in defining the S-matrix. Wave packets are used in most standard textbook treatments to construct particle states and remove certain singularities in the definition of the S-matrix. We show that this method does not completely work in curved spacetimes first in a general way and then by way of a specific model. We also discuss related effects and suggest a method for doing curved spacetime scattering calculations. Our conclusion is that the most general method for scattering calculations in curved spacetimes requires the use of wave packets, which are typically absent in the literature.}


\maketitle
\tableofcontents
\section{Introduction}\label{sec:intro}

Quantum field theory in curved spacetime is by now a well-established subject with several textbooks devoted to it (e.g. \cite{Birrell1982,Mukhanov2007}). Classic uses of the formalism include the study of particles near black holes \cite{Meda2021a,Zilberman2021,Taylor2021}, as well as scattering calculations in cosmological spacetimes \cite{Lankinen2017,Lankinen2018a,Lankinen2018b,Audretsch1985,Audretsch1987,Birrell1979,Birrell1979b}. Plenty of attention has also been devoted to the mathematical structure of the formalism \cite{Fewster2020,Fewster2021,Hollands2001,Hollands2002,Hollands2010,Wald1994,Wald1979S,Rejzner2021}.  

Our interest in this paper is scattering calculations, and more specifically the definition of the S-matrix. Specifically, we will look at the effect of wave packets on S-matrix calculations. The work of Ishikawa and others shows that finite time effects related to wave packets happen even in the case of flat spacetimes \cite{Ishikawa2005,Ishikawa2013,Ishikawa2014,Ishikawa2015,Ishikawa2016,Ishikawa2018,edery_wave_2021}. There is reason to believe that these effects are exacerbated when we are dealing with a curved spacetime. In this article we will show that the S-matrix has to be modified even in cases where the time span is infinite; a similar result in flat spacetime is due to Edery \cite{edery_wave_2021}.

The reason for our interest in the S-matrix, aside from the fact that it is a formally important part of any quantum field theory, is that these modifications to the S-matrix may lead to modified cross sections in cosmological processes. We will show that certain particle decay rates in e.g. cosmological spacetimes may be slightly modified due to wave packet corrections to the S-matrix. The spacetime need not even be curved; curved spacetime arguments apply in flat spacetimes with boundaries and any other scenarios in which the spacetime is not maximally symmetric (see e.g. references \cite{Balian_Duplantier_1977,Balian_Duplantier_1978,Klimchitskaya_Mohideen_Mostepanenko_2009,Brown_Maclay_1969,Dowker_Kennedy_1978} for the Casimir effect). Wave packets do not matter in flat space times with typical boundary conditions, where it is reasonable to discard their effects in most scenarios.

We will first provide a heuristic argument for why the S-matrix depends on the wave packet shape in curved spacetimes. We will then give a general formal argument for what this modification may look like and what its origin is. Finally, we will provide a solution to a novel spacetime with unusual boundary conditions that we use to illustrate these points and discuss the implications for scattering calculations in curved spacetimes.

\section{Wave packets and the S-matrix}\label{sec:arguments}
\subsection{A heuristic view}

In the standard Minkowski space treatment, the S-matrix is defined using wave packets. After a series of assumptions, the dependence of the S-matrix on the particular details of the wave packets finally vanishes. In the process, some of the singularities in the S-matrix are removed.  The use of sharply defined momentum states turns out to be a good approximation. Let us review the assumptions that go in to this definition.

\begin{enumerate}
    \item The region the theory is defined in is $(t,\mathbf{x})\in \mathbb{R}\times \mathbb{R}^3$.
    \item The states constructed using the wave packet, of the type $\int \Tilde{dk} \phi_P (k) a^\dagger _k |0\rangle$, have a recognizable central value $P$ around which the wave packet is peaked.
    \item Poincare invariance with resultant momentum and energy conservation.
\end{enumerate}

Unfortunately assumptions (1) and (3) are not necessarily true in curved space times. In arbitrary spacetimes momentum and energy are certainly not conserved and this includes cosmologically interesting metrics. Assumption (1) is also not always satisfied, because the system may be tightly constrained such that the fields are limited to some finite interval. Even assumption (2) contains subtleties which we will discuss in the following section. It has been shown that even in flat spacetimes finite time intervals may cause extra terms to appear in scattering calculations \cite{Ishikawa2005,Ishikawa2013,Ishikawa2014,Ishikawa2015,Ishikawa2016,Ishikawa2018}. This is because the wave packets overlap at finite times, and there is a heuristic reason to believe this correction would be even larger in curved spacetimes: the lack of momentum and energy conservation might cause e.g. a massive particle to decay in to two particles that travel 
in the same direction, thus maintaining an overlap between wave packets for a longer time.

In infinite time intervals, we still have a wave packet correction in curved spacetimes. Mathematically, this occurs because the lack of conservation laws eliminates the possibility to integrate out the wave packet dependence. Conceptually, the issue is that due to the lack of symmetry in the theory the momentum distributions of e.g. two particles may not be "separated enough"\ for interference effects to disappear. Investigating this correction in more detail is our aim in this paper. We will now illustrate it formally.

\subsection{A formal view}
From this point on, we will use the normalization
\begin{align}
    [a_k,a^\dagger_{p}] = 2\pi 2E_k \delta (k-p).
\end{align}
We will follow the standard textbook argument by Peskin and Schroeder \cite{Peskin1995}, restricting ourselves to 1+1 dimensions and two in-state particles with no loss of generality. The wave packets are constructed with
\begin{align}
    |\phi _P \rangle = \int \frac{dk}{\sqrt{2\pi 2E_k}}\phi (k)|k\rangle = \int \Tilde{dk}\phi_P (k)|k\rangle 
\end{align}
where $|k\rangle = a^\dagger _k|0\rangle$ and the states are normalized as 
\begin{align}
    \int dk |\phi _P (k)|^2 = 1.
\end{align}
Suppose we have two in-state particles and some number of out-state particles of momenta $\{p_f\}$. We use wave packets for the in-state only for reasons of simplicity, but they could equally well be added to the out-state. Then the transition probability is
\begin{align}
    \mathcal{P} = \prod _{f} \frac{dp_f}{2\pi 2E_{p_f}}|\langle \text{out},p_1..p_n|\phi _{P_1} \phi _{P_2},\text{in}\rangle|^2.
\end{align}
In terms of the wave packets, the squared amplitude is given by
\begin{align}
    |\langle \text{out},p_1..p_n|\phi _{P_1} \phi _{P_2},\text{in}\rangle|^2=&\int \Tilde{dk_1}\Tilde{dk_1'}\Tilde{dk_2}\Tilde{dk_2'}\phi _{P_1}(k_1)\phi _{P_2}(k_2) \phi _{P_1}(k_1')\phi _{P_2}(k_2')\nonumber\\
    &\times\langle \text{out},\{p_f\}|k_1k_2,\text{in}\rangle\langle \text{out},\{p_f\}|k_1'k_2',\text{in}\rangle .
\end{align}
If the spacetime were flat, we would now write the reduced matrix element as
\begin{align}
    \langle \text{out},\{p_f\}|\{k\},\text{in}\rangle = i2\pi \delta^{(2)} \bigg(\sum _f p_f - k-k'\bigg)\mathcal{M}(\{k\}\rightarrow \{p_f\}).
\end{align}
This is possible because of the conservation of momentum and energy. We then have two 2-dimensional delta functions $mathcal{P}$. We use the first one to get rid of two of the $k_i$ integrals. This yields in the standard treatment:
\begin{align}
    |\langle \text{out},p_1..p_n|\phi _{P_1} \phi _{P_2},\text{in}\rangle|^2=\int &\frac{dk_1dk_2}{ 4E_{k_1}E_{k_2}|v_{k_1}-v_{k_2}| }|\phi_{P_1} (k_1)|^2|\phi_{P_2} (k_2)|^2\nonumber\\
    &\times\delta^{(2)} \bigg(\sum _f p_f - k_1-k_2\bigg)|\mathcal{M}(k_1k_2\rightarrow \{p_f\})|^2.
\end{align}
We would then assume that the wave packets $\phi _P$ are strongly peaked around $P_1$ and $P_2$, so that we can remove all the remaining smooth functions from the integral ($k_i \approx P_i$), leaving us with just an integral over the wave packets. Since those are normalized, the wave packet dependence integrates out, and we are left with 
\begin{align}
    \mathcal{P} = \prod _{f} \frac{dp_f}{2\pi 2E_{p_f}} \frac{\delta ^{(2)}(\sum _{f}p_f - P_1-P_2)|\mathcal{M}(P_1P_2\rightarrow \{p_f\})|^2}{4E_{P_1}E_{P_2}|v_{P_1}-v_{P_2}|}
\end{align}
with $v_p=p/E_p$.

The lack of conservation laws makes this reasoning unworkable in curved spacetimes. Let us suppose momentum is conserved but energy is not, as is the case in the cosmological Robertson-Walker spacetimes. In that case,
\begin{align}
    \langle \text{out},\{p_f\}|\{k_i\},\text{in}\rangle = i2\pi \delta \bigg(\sum _f p_f - \sum _i k_i\bigg)\mathcal{M}(kk'\rightarrow \{p_f\}).
\end{align}
leading to
\begin{align}
    &|\langle \text{out},k_1..k_n|\phi _{P_1} \phi _{P_2},\text{in}\rangle|^2=\int \Tilde{dk_1}\Tilde{dk_1'}\Tilde{dk_2}\phi _{P_1}(k_1)\phi _{P_2}(k_2) \phi _{P_1}(k_1')\phi _{P_2}\bigg(k_1'-\sum _f p_f\bigg)\nonumber\\
    &\times (2\pi)^2 \mathcal{M}(k_1k_2\rightarrow \{k_f\})\mathcal{M}((k_1'k_2'-\sum _f p_f)\rightarrow \{k_f\})\delta \bigg( \sum _f p_f-k_1-k_2 \bigg)
\end{align}
Clearly, the wave packets remain even after performing the integrations. If we now made the same assumptions as in the flat spacetime case (i.e. assuming sharply peaked wave packets), we get
\begin{align}
    \mathcal{P}_{\text{curved}} = &\prod _{f} \frac{dp_f}{2\pi 2E_{p_f}} \frac{\delta (\sum _{f}k_f - P_1-P_2)|\mathcal{M}(P_1P_2\rightarrow \{p_f\})|^2}{4E_{P_1}E_{P_2}}\nonumber\\
    &\times \int dk_1dk_2dk_1' \phi _{P_1}(k_1)\phi _{P_2}(k_2)\phi_{P_1}(k_1')\phi _{P_2}\bigg( k_1'-\sum _f k_f \bigg) \label{eq:pcurvedsharp}
\end{align}
where the wave packet dependence does not disappear with the standard tricks.

In the final step of removing the second delta function from the integral, we also have to assume that the packets are narrow enough that the detector cannot distinguish between a packet and its central value. That is, we have to assume that the overlap is insignificant enough for the assumption 
\begin{align}
    \delta \bigg(\sum _f k_f - k_1 - k_2\bigg) \approx \delta \bigg(\sum _f k_f - P_1 - P_2\bigg).
\end{align}
This assumes the overlap effect to be small and the wave packet to be very sharply peaked, but this may not be a good approximation. If we drop that assumption, then in the case of Gaussian wave packets we have
\begin{align}
    &\int dk_1dk_2dk_1' \phi _{P_1}(k_1)\phi _{P_2}\bigg(k_1-\sum _f k_f\bigg)\phi_{P_1}(k_1')\phi _{P_2}\bigg( k_1'-\sum _f k_f \bigg) \nonumber\\
    &\propto \exp \bigg(-\frac{(\sum_f k_f-P_1-P_2)^2}{4\sigma}\bigg).
\end{align}
We are then left with a Gaussian packet term that would have to be accounted for in the integration over external momenta. It is not a priori clear that it can be ignored; at any rate, due to the lack of conservation laws, we are left with a wave packet dependence that has to be handled in some way.

The standard assumptions therefore lead to a wave packet dependence of the transition probability which remains even when taking limits of e.g. the width parameter $\sigma$. Given that the calculation itself gives us no reason to prefer one wave packet over another, this is a sure sign that one or more of the approximations made are inappropriate. Sharp momentum state calculations have nevertheless been used in the literature: references \cite{Lankinen2017,Lankinen2018a,Lankinen2018b,Audretsch1985,Audretsch1987,Birrell1979b} use the flat spacetime formula when calculating cross sections. This is reasonable enough because e.g. in the Robertson-Walker metric, the spatial part of the spacetime is maximally symmetric and there are no boundaries, so we should expect packet effects to be small.

In flat spacetime, we might be justified in assuming that the wave packets are Gaussian in the rest frame of one species of particle, since the particle momenta created for an experiment would likely be normally distributed. We might then perform a Lorentz boost to find the distribution in another frame, as done by Edery \cite{edery_wave_2021}. However, this approach does not work in curved spacetimes. First of all, QFT in curved spacetime is not Lorentz invariant in the first place, so Lorentz boosts are not the correct invariant transformation between frames. Conformally coupled massless particles are also useful indicators of scattering processes as they cannot be generated by the spacetime \cite{Audretsch1985,Audretsch1987} -- and there is obviously no Lorentz boost in to a frame moving at the speed of light even in flat spacetime. 

We are left with three ways to perform an S-matrix calculation: first, we could discard most of the assumptions of the flat spacetime calculation. Second, we could keep most of them -- such as the wave packets being peaked around a particular value -- but take in to account the reduced symmetry of the spacetime. Finally, we could hope the flat spacetime formula works approximately and ignore the wave packets. We will now study a model with all three approaches.
\section{Scattering in curved spacetimes: three ways}\label{sec:sec3}
Let us clarify the situation by specifying a concrete model. The basic principles apply to other metrics rather generally. We will use a spacetime restricted to $(t,x)\in \mathbb{R}\times \mathbb{R}^+$ with the metric
\begin{align}
    g_{\mu \nu} = 2ax  \eta _{\mu\nu},\quad a>0,\quad x\geq 0. \label{eq:metric}
\end{align}
The free Lagrangian is
\begin{align}
    \mathcal{L} = \frac{1}{2}\sqrt{-g}\bigg[\partial ^\mu \phi \partial _\mu - m^2 \phi \bigg]
\end{align}
for the massive particle $\phi$, and the corresponding massless version for $\psi$. This metric was chosen because it has both a boundary and breaks a conservation law, the importance of which shall be discussed presently. It also has the virtue of being solvable. The solutions are
\begin{align}
    \psi (x) &= \int \frac{dk}{\sqrt{2\pi} 2E_k} \sqrt{\frac{2}{\pi}}\sin (kx) \bigg(a_k^\dagger e^{i\omega t} + a_k e^{-i\omega _t}\bigg),\\
    \phi (x) &= \sum _i \frac{\mathcal{N}}{\sqrt{\omega _i}}\text{Ai}(f(x,\omega _i))\bigg[ e^{i\omega _i t}a^\dagger _p + e^{-i\omega _i t}a_p \bigg].
\end{align}
for the massless ($\psi$) and massive ($\phi$) particles respectively. The boundary condition is $\psi (0) = \phi (0) = 0$ and
\begin{align}
    \mathcal{N} &= \sqrt{\frac{(2am^2)^{1/3}}{\text{Ai}'(\zeta _i)}},\\
    f(x,\omega _p) &= \frac{-\omega _p + 2am^2x}{2^{2/3}(am^2)^{2/3}}.
\end{align}
The spectrum of the massive field is discrete: $\omega _i^2 = - {2^{2/3}(am^2)^{2/3}}\zeta _i $. The $\zeta _i$ are Airy function zeroes, which are all negative. The spectrum of the massless solution is also limited to $k\in [0,\infty]$ (equivalently $k\in [-\infty ,0]$) since those are the only linearly independent solutions.

We will use the interaction
\begin{align}
    \mathcal{L}_I = -\lambda  \psi (x)^2\phi (x)\sqrt{-g}.
\end{align}

We will be calculating the creation of two massless particles by a single massive particle. We are interested in the transition probability $\mathcal{P} = |\langle \text{out}, p_1p_2|p,\text{in}\rangle|^2$. For sharp momentum states, we have
\begin{align}
    &2\pi \delta (E_p-E_{p_1}-E_{p_2}) \mathcal{M}(p\rightarrow p_1p_2) = \langle \text{out},p_1p_2|p,\text{in}\rangle \nonumber\\
    &=-\lambda \frac{\mathcal{N}}{\sqrt{\omega _p}}(2\pi)^2 \delta (E_p-E_{p_1}-E_{p_2})\int dx \sqrt{-g} \frac{2}{\pi}\sin (p_1x) \sin (p_2x)\text{Ai}(f(x,\omega _p)).\label{eq:mtransition}
\end{align}
\subsection{General wave packets }\label{sec:dull}
We will first calculate the transition probability without making any of the standard approximations. We neither assume the conservation of momentum and energy nor that the wave packets are peaked around $P_1$ and $P_2$. We start from
\begin{align}
    &\mathcal{P}_{\text{general}}(P_1,P_2)\nonumber\\
    &=|\langle \text{out},\phi _{P_1}\phi _{P_2}|p,\text{in}\rangle|^2 \\
    &= \bigg|\int \frac{dk_1dk_2}{(2\pi)2\sqrt{k_1k_2}}\phi_{P_1} (k_1)\phi_{P_2}(k_2)2\pi \delta (E_p-E_{k_1}-E_{k_2}) \mathcal{M}(p\rightarrow p_1p_2)\bigg|^2\\
    &= \int \frac{dk_1dk_2dk_1'dk_2'}{4\sqrt{k_1k_2k_1'k_2'}}\bigg[\phi_{P_1} (k_1)\phi_{P_2}(k_2)\phi_{P_1} (k_1')^*\phi_{P_2}(k_2')^*\nonumber\\
    &\quad\times \delta (E_p-E_{k_1}-E_{k_2})\delta (E_p-E_{k_1'}-E_{k_2'})\mathcal{M}(p\rightarrow k_1k_2)\mathcal{M}^*(p\rightarrow k_1'k_2')\bigg]
\end{align}
In our case, since $k_i>0$, we have $E_{k_i}=k_i$. Therefore
\begin{align}
    &\mathcal{P}_{\text{general}}(P_1,P_2)\nonumber\\
    &=\int \frac{dk_2dk_2'}{4\sqrt{(E_p-k_2)k_2(E_p-k_2')k_2'}}\bigg[\phi_{P_1} (E_p-k_2)\phi_{P_2}(k_2)\phi_{P_1} (E_p-k_2')^*\phi_{P_2}(k_2')^*\nonumber\\
    &\quad\times \mathcal{M}(p\rightarrow (E_p-k_2)k_2)\mathcal{M}^*(p\rightarrow (E_p-k_2')k_2')\bigg]. \label{eq:dullstuff}
\end{align}
with
\begin{align}
    \mathcal{M}(p\rightarrow p_1 p_2)=\lambda \frac{\mathcal{N}}{\sqrt{\omega _p}}\int dx \sqrt{-g}\frac{2}{\pi} \sin (p_1x) \sin (p_2x)\text{Ai}(f(x,\omega _p)).
\end{align}
The $k_i$-integrals are now in the region $k_i\in [0,E_p]$ due to the conservation of energy.

This should be the most generic form of a scattering calculation in this spacetime, and analogously using the same steps it would be the most generic form of a scattering calculation in any spacetime for which a quantum field theory is well defined (see e.g. \cite{Wald1994} for the restrictions on the spacetime). Unfortunately, it is often difficult enough to calculate integrals over the matrix elements $\mathcal{M}$, and adding wave packets to the mix makes the situation worse. In the case of our metric, for instance, even numerical integration of \eqref{eq:dullstuff} is numerically unstable with standard algorithms\footnote{As implemented in Mathematica v. 14.0.}.

Practical necessity therefore forces us to approximate.
\subsection{Sharp wave packets}\label{sec:sharp}
We now add assumptions used in flat spacetime. Specializing eq. \eqref{eq:pcurvedsharp} to our process, we have
\begin{align}
    \mathcal{P}_{\text{sharp}}(P_1,P_2) &=  \int \frac{dk_1dk_2dk_1'dk_2'}{4\sqrt{k_1k_2k_1'k_2'}}\phi _{P_1}(k_1)\phi _{P_2}(k_2)\phi _{P_1}(k_1')\phi _{P_2}(k_2')\nonumber\\
    &\quad \times \mathcal{M}(p\rightarrow k_1k_2)\mathcal{M}^*(p\rightarrow k_1'k_2')\delta(E_p-E_{k_1}-E_{k_2})\delta(E_p-E_{k_1'}-E_{k_2'})
\end{align}
We use the delta functions and make the assumption that the functions inside the integral are sharply peaked around $P_1,P_2$ so that we can pull them out of it. This yields immediately
\begin{align}
    \mathcal{P}_{\text{sharp}}(P_1,P_2) &= \frac{|\mathcal{M}(p\rightarrow P_1P_2)|^2}{4E_{P_1}E_{P_2}}\nonumber\\
    &\quad\times\int dk_1dk_2' \phi _{P_1}(k_1)\phi _{P_2}(E_p-k_1)\phi _{P_1}(E_p-k_2')\phi _{P_2}(k_2')\nonumber\\
    &= \frac{|\mathcal{M}(p\rightarrow P_1P_2)|^2}{4E_{P_1}E_{P_2}} I_{P_1,P_2}\label{eq:sharppack}
\end{align}
The first part of \eqref{eq:sharppack} is basically the flat spacetime result with a reduced spacetime symmetry; the overlap integral $I_{P_1P_2}$ provides a slight correction. We have used the delta functions arising from the conservation laws in arriving at \eqref{eq:sharppack}, rather than taking one of the delta functions out of the integral, as is done in the flat spacetime case. Taking the delta function out of the integral amounts to the approximation that the detector is not sensitive enough to pick up interference effects \cite{Peskin1995}, which is contrary to our premise.

For instance, we could use a truncated Gaussian packet normalized on the interval $k\in [0,\infty]$, so that
\begin{align}
    \phi_P (k) = \frac{\sqrt{2} e^{-\frac{(k-P)^2}{2 \sigma}}}{\sqrt[4]{\pi } \sqrt{\sqrt{\sigma} \left(\text{erf}\left(\frac{P}{\sqrt{\sigma}}\right)+1\right)}}.
\end{align}
where erf is the error function. In that case, we would get for \eqref{eq:sharppack} 
\begin{align}
    \mathcal{P}_{\text{sharp}}(P_1,&P_2)\overset{\sigma\ \text{small}}{\approx} \frac{|\mathcal{M}(p\rightarrow P_1P_2)|^2}{4E_{P_1}E_{P_2}}\frac{e^{-\frac{\left(-\omega _p+P_1+P_2\right){}^2}{2 \sigma }} \left(\text{erf}\left(\frac{\omega _p+P_1-P_2}{2 \sqrt{\sigma }}\right)+1\right){}^2}{\left(\text{erf}\left(\frac{P_1}{\sqrt{\sigma }}\right)+1\right) \left(\text{erf}\left(\frac{P_2}{\sqrt{\sigma }}\right)+1\right)}. \label{eq:psharp}
\end{align}
Note that since we have used our delta functions from the conservation laws already, $\omega _p$ is only approximately equal to $P_1+P_2$. This expression does not have a well-defined limit $\sigma \rightarrow 0$.

In this manner, we can get rid of the wave packet integrals as we can in flat spacetime and the problem has become slightly more tractable once again. Yet we still have a complicated-looking wave packet term. If we wanted to now integrate over the external momenta $P_1$ and $P_2$, we would likely encounter practical difficulties in obtaining an analytical result.
\subsection{No wave packets}
The standard method used in the literature \cite{Lankinen2017,Lankinen2018a,Lankinen2018b,Audretsch1985,Audretsch1987,Birrell1979b} has been to simply ignore wave packets and the foregoing discussion. We, in effect, assume that the spacetime is "flat enough"\ at the length and time scale of the scattering process. This is a difficult assumption to directly quantify. To proceed, we would have to set up our theory in a box to make the spectrum discrete. In our case, it is sufficient to limit the time integration to some interval $t\in [0,T]$ since the momentum is not conserved. Then
\begin{align}
    \psi (x) = \frac{1}{V}\sum _{k_i}\bigg[ f_{k_i}(x)^*a^\dagger _{k_i} + f_{k_i}(x)a _{k_i}\bigg].
\end{align}
When we run in to a double delta function while calculating $\mathcal{P}$ and want to take the limit of $V\rightarrow \infty $, we use the standard trick on the limit:
\begin{align}
    \delta _{pkk'}\delta _{pkk'} &\rightarrow (2\pi )^2\delta (\omega _p-\omega_k-\omega_{k'})\delta (\omega _p-\omega_k-\omega_{k'})\\
    &=2\pi \delta (\omega _p-\omega_k-\omega_{k'})\int dt e^{i(\omega _p-\omega_k-\omega_{k'})t}\\
    &=2\pi \delta (\omega _p-\omega_k-\omega_{k'})T.
\end{align}
Adding this to the previous assumptions, we end up with
\begin{align}
    \mathcal{P}_{\text{no packet}}(P_1,P_2) =dP_1dP_2\frac{\delta (\omega _p-P_1-P_2)|\mathcal{M}(p\rightarrow P_1P_2)|^2}{4E_{P_1}E_{P_2}}T. \label{eq:pnopacket}
\end{align}
The total probability now depends on interaction time: over an infinite time interval, the decay process will eventually happen. In the wave packet calculation this factor is suppressed because of the assumptions we have implicitly made about the wave packet evolution. In a full wave packet Fock basis, we would likely find a similar time dependence, as the massive particle's wave packet would likely spread over time and thus have perpetual overlap with the out-state massless particles.

What are the differences between \eqref{eq:pnopacket} and \eqref{eq:psharp}? Note that the packetless probability is defined only when $\omega _p = P_1+P_2$. For the sharp wave packet, this is not the case; since to arrive at \eqref{eq:psharp} we have already used the delta functions arising from conservation laws, $\omega _p \neq P_1+P_2$. The wave packet version therefore includes an exponential fudge factor in the momentum distribution.

It is worth pointing out that we could also dsicretize the spacetime to tame the singularities. However, this sort of discretization is especially problematic in curved spacetimes, as some geometrical features would be lost -- it is not usually possible to discretize objects from differential geometry without losing some of their properties, as discussed in ref. \cite{crane_glimpse_2017}.
\section{Discussion}
In the previous section we calculated the transition probability of a massive particle decaying in to two massless particles in three different ways. Each leads to a manifestly different result. This does not happen in a flat spacetime, as it can be shown in a general way that the S-matrix will never depend on the shape of the wave packet as long as the requisite assumptions are satisfied. 

These results can be applied in a relatively straightforward way to more complicated examples in, for example, 4 spacetime dimensions. The wave packet corrections look essentially the same: having computed a squared amplitude $|\langle \text{out},\{ p \} | \{k\},\text{in}\rangle |^2$, the sharp wave packet version simply requires an addition of the overlap integral:
\begin{align}
    |\langle \text{out},\{ p \} | \{k\},\text{in}\rangle|^2\rightarrow |\langle\text{out}, \{ p \} | \{k\},\text{in}\rangle|^2 I_{\{ p\}\{k\}}
\end{align}
with $I_{\{ p\}\{k\}}$ computed as in \ref{sec:sharp}.

We emphasize that the foregoing results depend only on the lack of symmetry; the physical reason for this is not relevant. Therefore the results should apply to systems with symmetry-breaking boundary conditions -- we deliberately introduced such a condition in our calculation in sec. \ref{sec:sec3}. Boundary effects in QFT have a rich history in the literature.  The textbook of Birrell and Davies contains a list of such phenomena \cite{Birrell1982}; of particular interest is the experimentally testable Casimir effect, see e.g. \cite{Klimchitskaya_Mohideen_Mostepanenko_2009} for a review.

Ishikawa et al. showed that in flat spacetimes, the wave packet remains approximately Gaussian throughout its evolution \cite{Ishikawa2018}. With a sufficiently large ("infinite") time interval for the calculation, the Gaussians become separated and can be treated as sharp. If the time interval is short, the wave packets are not yet far away from one another and there are overlap effects that depend on their shape, but these are predicted to be small for most processes, especially those happening in a laboratory on Earth. Some astronomical processes might be affected \cite{Ishikawa2015}.

Here, we have no such assurances. First of all, the evolution of a wave packet is not going to "stay Gaussian"\ in even an approximate sense. Even in our simple and rather pathological example, this is true for the massive particles in the metric \eqref{eq:metric}, and even massless particles experience boundary effects. This is a problem because, as we have seen, a dependence on the wave packet is left in the transition probability. How can we know that the distributions of momentum states are Gaussian for both in and out particles? As we noted previously, it might make sense to say we can create a Gaussian distribution in a resting laboratory frame, but when dealing with massless particles and a theory that is not Lorentz-invariant, we can no longer simply assume that the outgoing particle stream has a Gaussian distribution. Which packet should we choose? One possible method for constructing the packet might be to include a model of the detector in to the calculation as was done by Unruh in his famous paper \cite{unruh_notes_1976}; this might be an interesting research direction.

The correct wave packet might also not have a sharp limit as we would have in the textbook derivation of the transition probability; that depends on the distribution being Gaussian and sufficiently narrow. In the general case, the assumption of a narrow distribution might not be sensible at all, and this is the difference between the calculations in sections \ref{sec:dull} and \ref{sec:sharp}.

In our view, the most conceptually sound option is to bite the bullet and go with the most cumbersome method of \ref{sec:dull}. One can perform the calculation as far as possible without specifying the wave packet, and then attempt a case-by-case analysis of the spacetime and the interaction to see which form is correct. More convenient is the method of \ref{sec:sharp}, which only adds the assumption of a narrow distribution of momentum states. The main thrust of our argument is that in the general case, dismissing the wave packets is not justified, though it is approximately justified in cases of high symmetry (e.g. Robertson-Walker metric). This is why the calculations done in e.g. \cite{Lankinen2017,Lankinen2018a,Lankinen2018b,Audretsch1985,Audretsch1987,Birrell1979b} are at least approximately justified.

It might seem startling that the wave packet dependence does not disappear. However, this is not really surprising; many pleasant properties of flat spacetime QFT disappear in curved spacetimes due to the lack of symmetry in the theory. As was shown by Edery \cite{edery_wave_2021}, effects similar to those in this paper may be relevant even in the Minkowskian case. The issues are simply exacerbated in curved spacetimes.

It would be interesting to perform the finite-time analysis of Ishikawa et al. on curved spacetimes. This, however, requires writing the theory in a wave packet picture, and it is not easy to find a set of wave packets that span the Fock space of a curved spacetime QFT. Even if you can find such a set, the calculations would again be cumbersome and likely less illuminating than the flat spacetime case.

In summary, we investigated three different approaches for doing scattering calculations in curved spacetime. Despite the prevalence of the packetless method in the literature, we think it might need modification: the next time you receive a parcel of particles, keep the packaging. It is the method of sec. \ref{sec:dull} which is the most conceptually sound, and \ref{sec:sharp} should also be reasonable as long as the momentum distribution is not too spread out. We have also provided a solution to a spacetime with $g=2ax\eta $, which might be useful for e.g. conceptual purposes. 

In future work, it would be interesting to investigate how different spacetimes affect the momentum distributions of particles prepared by an experimenter or measured by a detector.

\bibliography{sn-bibliography}

\end{document}